\documentclass[a4paper,onecolumn,11pt]{quantumarticle}
\pdfoutput=1
\usepackage[utf8]{inputenc}
\usepackage[english]{babel}
\usepackage[T1]{fontenc}
\usepackage{amsmath}
\usepackage{hyperref}
\usepackage{comment}
\usepackage[backend=biber,style=ieee,sorting=none]{biblatex}
\usepackage[bblfile=main]{biblatex-readbbl}
\usepackage{xurl}
\usepackage{CJKutf8}

\begin{document}
\author{W4Q:}
\affiliation{$^{*}$ \textbf{Authors by alphabetic first name order}}
\affil[*]{\textbf{Corresponding authors}: Almut Beige (a.beige@leeds.ac.uk), Anna Sanpera (Anna.Sanpera@uab.cat), Christiane Koch (christiane.koch@fu-berlin.de), Ivette Fuentes  ( I.Fuentes-Guridi@soton.ac.uk), Marilu Chiofalo (marilu.chiofalo@unipi.it), Roberta Zambrini (roberta@ifisc.uib-csic.es),\\
Valentina Parigi (valentina.parigi@lkb.upmc.fr), Sabrina Maniscalo (sabrina.maniscalco@helsinki.fi)\\}
\author{Almut Beige}
\affiliation{The School of Physics and Astronomy, University of Leeds, United Kingdom}
\author{Ana Predojevi\'{c}}
\affiliation{Department of Physics, Stockholm University, Sweden }
\author{Anja Metelmann}
\affiliation{Institute for Theory of Condensed Matter, Karlsruhe Institute of Technology,  Germany}
\affiliation{Institute for Quantum Materials and Technology, Karlsruhe Institute of Technology, Germany}
\affiliation{ISIS, Université de Strasbourg, France}

\author{Anna Sanpera}
\affiliation{Informació i Fenòmens Quàntics. Departament de Física, Universitat Autònoma de Barcelona, Spain}
\affiliation{ICREA, Institució Catalana de Recerca i Estudis Avançats, Spain}

\author{Chiara Macchiavello}
\affiliation{Dipartimento di Fisica, Università degli Studi di Pavia, Italy}
\affiliation{INFN Sezione di Pavia, Italy}

\author{Christiane P. Koch}
\affiliation{Dahlem Center for Complex Quantum Systems and Fachbereich Physik, Freie Universität Berlin, Germany}

\author{Christine Silberhorn}
\affiliation{Integrated Quantum Optics Group, Department of Physics,
Institute for Photonic Quantum Systems (PhoQS),
Paderborn University,
Germany
}

\author{Costanza Toninelli}
\affiliation{National Institute of Optics [(Consiglio Nazionale delle Ricerche CNR)–INO)],
Sesto Fiorentino, Italy}
\affiliation{European Laboratory for Non-Linear Spectroscopy (LENS), 
Sesto Fiorentino, 
Italy}

\author{Dagmar Bruß}
\affiliation{Institut für Theoretische Physik III, Heinrich-Heine-Universität Düsseldorf, 
Germany}

\author{Elisa Ercolessi}
\affiliation {Dipartimento di Fisica e Astronomia, University of Bologna, Italy}
\affiliation{INFN Sezione di Bologna, Italy}

\author{Elisabetta Paladino}
\affiliation{Dipartimento di Fisica e Astronomia “Ettore Majorana”, Università di Catania, Italy}
\affiliation{INFN, Sezione di Catania, Italy}
\affiliation{CNR-IMM, UoS Università, Catania, Italy}

\author{Francesca Ferlaino}
\affiliation{Institut für Quantenoptik und Quanteninformation, Österreichische Akademie der Wissenschaften, Innsbruck, Austria}
\affiliation{Universität Innsbruck, Institut für Experimentalphysik, 
, Austria}

\author{Giulia Ferrini}
\affiliation{Wallenberg Centre for Quantum Technology, Department of Microtechnology and Nanoscience, Chalmers University of Technology , 
Göteborg, Sweden}

\author{Gloria Platero}
\affiliation{Theoretical Condensed Matter Departament, Materials Science Institute of Madrid (CSIC), Spain}

\author{Ivette Fuentes}
\affiliation{School of Physics and Astronomy, University of Southampton, United Kingdom}

\author{Kae Nemoto}
\affiliation{Okinawa Institute of Science and Technology Graduate University, Japan}
\affiliation{National Institute of Informatics, 
Tokyo, Japan}

\author{Leticia Tarruell}
\affiliation{ICFO, Institut de Ciències Fotòniques, The Barcelona Institute of Science and Technology, Castelldefels, Barcelona, Spain}
\affiliation{ICREA, Institució Catalana de Recerca i Estudis Avançats, Spain}

\author{Maria Bondani}
\affiliation{CNR-Institute for Photonics and Nanotechnologies, Como, Italy}

\author{Marilu Chiofalo}
\affiliation{Department of Physics “Enrico Fermi”, University of Pisa, and INFN-Sezione di Pisa, Italy}

\author{Marisa Pons}
\affiliation{EHU Quantum Center, Universidad del País Vasco, UPV/EHU, Leioa, Spain}
\affiliation{Departamento de Física Aplicada, Universidad del País Vasco, UPV/EHU, Bilbao, Spain}

\author{Milena D’Angelo}
\affiliation{Dipartimento Interuniversitario di Fisica, Università degli studi di Bari, Italy}
\affiliation{INFN, Sezione di Bari, Italy}

\author{Mio Murao}
\affiliation{Department of Physics, Graduate School of Science, The University of Tokyo, 
Japan}

\author{Nicole Fabbri}
\affiliation{National Institute of Optics [(Consiglio Nazionale delle Ricerche CNR)–INO)],
Sesto Fiorentino, Italy}
\affiliation{European Laboratory for Non-Linear Spectroscopy (LENS), 
Sesto Fiorentino, 
Italy}

\author{Paola Verrucchi}
\affiliation{ Dipartimento di Fisica e Astronomia, Università di Firenze,
Italy}
\affiliation{Istituto dei Sistemi Complessi, Consiglio Nazionale delle Ricerche, 
 Sesto Fiorentino , Italy}
\affiliation{Istituto Nazionale di Fisica Nucleare, Sezione di Firenze, 
 Sesto Fiorentino, Italy}

\author{Pascale Senellart-Mardon}
\affiliation{Université Paris-Saclay, CNRS, Centre de Nanosciences et de Nanotechnologies, Palaiseau, France}

\author{Roberta Citro}
\affiliation{Dipartimento di Fisica "E.R.Caianiello", University of Salerno, Italy}

\author{Roberta Zambrini}
\affiliation{Institute for Cross-Disciplinary Physics and Complex Systems (IFISC) UIB-CSIC, Campus Universitat Illes Balears, Palma de Mallorca, Spain}

\author{Rosario González-Férez}
\affiliation{Departamento de Física Atómica, Molecular y Nuclear, Universidad de Granada, Spain}
\affiliation{Instituto Carlos I de Física Teórica y Computacional, Universidad de Granada, Spain}

\author{Sabrina Maniscalco}
\affiliation{University of Helsinki, Finland}

\author{Susana Huelga}
\affiliation{Institut für Theoretische Physik und IQST, 
Universität Ulm, Germany}

\author{Tanja Mehlstäubler}
\affiliation{Physikalisch-Technische Bundesanstalt, 
Braunschweig, Germany}
\affiliation{Institute for Quantum Optics \& Laboratory for Nano- and Quantum Engineering, Universität Hannover, 
 Germany}

\author{Valentina Parigi}
\affiliation{Laboratoire Kastler Brossel, Sorbonne Université, CNRS, ENS-Université PSL, Collège de France Paris, 
France}

\author{Ver\'onica Ahufinger}
\affiliation{Departament de Física, Universitat Autònoma de Barcelona, 
Bellaterra, Spain}

\title{Women for Quantum - Manifesto of Values}

\begin{abstract}
Data show that the presence of women in quantum science is affected by a number of detriments and their percentage decreases even further for higher positions. Beyond data, from our shared personal experiences as female tenured quantum physics professors, we believe that the current model of scientific leadership, funding, and authority fails to represent many of us. It is time for a real change that calls for a different kind of force and for the participation of everyone. Women for quantum calls for a joint effort and aims with this initiative to contribute to such a transformation.
\end{abstract}

\maketitle

\section*{Introduction: Who we are}
Women for Quantum (W4Q) is a group of female physics professors currently mostly based in Europe but also in Japan, working in the field of AMO physics (quantum optics, atomic and molecular), quantum many-body physics, and quantum information. All of us hold tenured positions and have more than 10 years of professional experience post-PhD.
We have observed numerous initiatives aimed at improving gender balance in our field and making our professional environment more welcoming to diversity.
However, we find that most of these initiatives are ineffective in achieving these goals. Several facts, for the EU28 countries, 
confirm our observations as seen in \cite{Shefigures2021,dpg,IOP,dataspeak,Atominnen,Japan}:

\begin{itemize}
\item In the broader field of Natural Sciences, including biology, women account for
less than 22\% of full professors or equivalent research positions, ( 6 \% in Japan) despite
comprising more than 50\% at the university entry level.

\item In physics, the situation is even direr. Available data show that, for example, in
2021 Germany had 12\% full professors in physics, while the UK had the same
percentage in 2019, and Spain had 14\% in 2020. 
\item The phenomenon of women abandoning a research career, thus widening the
gender gap at every career stage (the “leaky pipeline”), is widespread.
\item Participation in research funding remains very low for women.
\item Women generally shoulder an excessive burden of community service
(committees, evaluations, etc.) compared to their male colleagues.
\item The gender pay gap in academia remains high in many countries.

\end{itemize}

These facts, combined with our shared personal experiences of uncomfortable environments for women \cite{Oreal}, have prompted us to reconsider the current
state of research practices: Who defines academic authority and how is it done? What is the present model of scientific leadership in general, and in our community in particular? How is funding of research organized? And why are research careers
excessively competitive to the extent of often hindering scientific progress?

We firmly believe that the current model of scientific leadership, funding, and authority fails to represent many of us. We are convinced that it is also detrimental to our colleagues, women, men, and non-binary persons.

It is time for a real change that calls for the participation of everyone. We aim with this initiative to be a seed for such a change.

\section*{What this Manifesto of Values is about}

This Manifesto outlines the values and goals we identify with. By sharing these values, we aim to initiate a dialogue and trigger new paths of doing research, thereby advancing the community as a whole. We trust that these values resonate with other members of our community, particularly those from underrepresented groups.

\subsection*{Values and goals}
\begin{itemize}
\item We value acting as a community, enriched by a genuine culture of sharing and
collaboration.
\item We value trust, honesty and integrity.
\item We value critical, curiosity-driven, and creative thinking.
\item We value diversity, and believe in empowering others.
\item We value the freedom to express opinions or ask questions without fear of judgment.
\item We value respect and kindness in discussions and communication, regardless of
hierarchy or role, in lieu of aggressive attitude.
\item We value respectful language and imagery in communicating scientific results or
lecturing.
\item We value using words and images that have the power to transform reality, creating an
inclusive and non-discriminatory, e.g. non-gender-biased, scientific community.
\item We value the respect for the space of the body and the comfort of each person.
\end{itemize}

Building on these values, we aim at

\begin{itemize}
    \item Taking an active role in transforming the way quantum science is done and valued,
both for ourselves and for future generations of (quantum) scientists;
\item Creating an open and welcoming space for female scientists;
\item Building a participatory, inclusive and supportive scientific community, where
teamwork is fostered: a community built on the concept of authority originating from
the Latin ‘augere’, i. e., focused on ‘nurturing’ and ‘growing’;
\item Bringing women to the forefront, which also entails redistributing power. Power is the
ability to make changes, create new possibilities and participate in decision making.
Power requires access to the right resources, including finances and connections.
Currently, female scientists are under-represented in decision-making bodies and de-facto prevented from exerting considerable influence on decision-making processes – relegating them to the position of being the ‘female in the room’ instead of being recognized as scientists. We advocate for power and decision-making to be shared rather than concentrated. We advocate for decision making to be integrated across the entire organizational structure and in particular at high levels, prioritizing transparency and focusing on the benefit of the entire scientific community.
\item Counteracting gender-bias in the scientific community. Gender-bias, including unconscious, is a well-documented phenomenon that affects the recognition and access of female scientists to financial resources, promotions, publication in high impact journals, among other areas. Our goal is to raise awareness on this
phenomenon, and to enforce effective measures to address it in its various, even subtle, forms.
\item Freeing our community from microaggressions, harassment, and any other degrading behaviors and practices that result in making women invisible or even just uncomfortable. Reports from many universities consistently show that women are more exposed to such discrimination and practices than the average. Additionally too many of us have personal stories or know colleagues who have experienced serious
misconduct, damaging women's careers. This misconduct ranges from belittling, to sexist comments, to actual harassment. We seek to raise awareness across the
whole scientific community about the current situation and strive to put an end to it.
\item Fostering, empowering, and implementing a different approach to evaluating quality in science. We believe it is crucial to assess not only the scientific output but also the process and journey leading to it. Therefore, we advocate for measures that go beyond relying solely on numerical metrics like the h-index or citation counts, also acknowledging the existence of diverse career paths. In recruitment and funding
processes, we advocate for assessing capabilities in teamwork and valuing one’s ability to create a healthy scientific ecosystem, in addition to technical and
managerial abilities.
\end{itemize}

We aim at achieving a true change. The first step is to acknowledge the unsatisfactory current situation of women in quantum physics. Existing measures have not enough impact. The change we seek will benefit all under-represented communities as well as the quantum
ecosystem as a whole.

\section*{Acknowledgements}
\noindent 
We would like to thank the many colleagues from (quantum) physics and the 
quantum industry, in particular the (first) endorsers of our manifesto, 
for their support and discussions over the past years.


\printbibliography 

\section*{First endorsers:}

Dorit Aharonov, 	Hebrew University Jerusalem,	Israel	\\
Alessia	Allevi,	Università degli Studi dell’Insubria,	Italy	\\
Nina Amini,	Laboratoire Signaux et Systèmes, Centrale Supélec, Université Paris-Saclay, France	\\
Janet Anders,	Universität Potsdam,	Germany	\\
Marta Anguiano Millán	, Universidad de Granada, Spain	\\
Sonia Antoranz Contera, University of Oxford,	UK	\\
Myrto Arapinis,	University of Edinburgh, UK	\\
Natalia	Ares, University of Oxford,	UK	\\
Alexia	Auffèves, CNRS MajuLab, CQT,	Singapore	\\
Ticijana Ban,	Institute of Physics, Zagreb,	Croatia	\\
Mari Carmen Bañuls, Max-Planck-Institut für Quantenoptik, Garching,	Germany	\\
Stefanie Barz, Universität Stuttgart,	Germany	\\
Leni Bascones, Instituto de Ciencia de Materiales de Madrid, CSIC, Spain	\\
Luisa Bausá, Universidad Autónoma de Madrid, Spain\\
Karine Beauchard, Institut de Recherche Mathématique de Rennes (IRMAR), ENS Rennes, France	\\
Silvia	Bergamini,	The Open University,	UK	\\
Blanca	Biel,	Universidad de Granada,	Spain	\\
Kerstin	Borras,	DESY and RWTH Aachen,	Germany	\\
Isabelle Bouchoule, Institut d'Optique, Palaiseau,	France	\\
Nadia	Bouloufa,	Laboratoire Aimé Cotton,	France	\\
Anne	Broadbent,	University of Ottawa,	Canada	\\
Natalia	Bruno,	CNR-INO and LENS, Sesto Fiorentino,	Italy	\\
Humeyra 	Caglayan , Tampere University,	Finland	\\
María José	Calderón, Instituto de Ciencia de Materiales de Madrid ICMM-CSIC,	Spain	\\
Francesca	Calegari, Center for Free-Electron Laser Science, DESY, and  University of Hamburg,	Germany	\\
María del Carmen Carrión Pérez,	Universidad de Granada,	Spain	\\
Lucia	Caspani,	Università degli Studi dell’Insubria,	Italy	\\
Hilda	Cerdeira,	Instituto de Física Teórica, UNESP, São Paulo,	Brasil	\\
Ana María	Cetto	,	Universidad Nacional Autónoma de México,	Mexico	\\
Caroline	Champenois,	PIIM, CNRS and Aix-Marseille University,	France	\\
Cecilia	Clementi,	Freie Universität Berlin,	Germany	\\
Maja	Colautti,	CNR-INO and LENS, Sesto Fiorentino,	Italy	\\
Audray	Cottet,	CNRS -LPENS-LPEM,	France	\\
Sarah	Croke,	University of Glasgow,	UK	\\
Otti	D'Huys,	Maastricht University,	Netherlands	\\
Irene	D’Amico,	University of York,	UK	\\
Virginia	D’Auria,	Université Côte d'Azur, INPHYNI,	France	\\
Nilanjana	Datta, 	University of Cambridge,	UK	\\
Sandra	De Jesus Raimundo,	University of Southampton,	UK	\\
Thereza Cristina	de Lacerda Paiva,	Universidade Federal do Rio de Janeiro,	Brazil	\\
Cristiane   de Morais Smith, University of Utrecht,	Netherlands	\\
Carla	De Morisson Faria,	University College London,	UK	\\
Inés	de Vega, IQM Quantum Computers,	Germany	\\
Krissia	de Zawadzki,	Universidade de São Paulo, São Carlos,	Brazil	\\
Eleni Diamanti,	Sorbonne Université, France	\\
Mina Doosti, University of Edinburgh, UK	\\
Claudia	Draxl, Humboldt Universität zu Berlin, Germany	\\
Anaïs Dréau,	Laboratoire Charles Coulomb - CNRS,	France	\\
Sara Ducci,	Université de Paris Cité, Laboratoire Matériaux et Phénomènes Quantiques,	France	\\
Deborah	Dultzin,	Universidad Nacional Autónoma de México,	Mexico	\\
Claudia	Eberlein, University of Loughborough, UK \\
Martina	Esposito,	CNR-SPIN Naples,	Italy	\\
Marta P. Estarellas,	Qilimanjaro Quantum Tech,	Spain	\\
Karin Everschor-Sitte, Universität Duisburg-Essen, Germany	\\
Margarida	Facao,	Universidade de Aveiro,	Portugal	\\
Anita Fadavi Roudsari, Chalmers University of Technology, Sweden	\\
Clotilde Fermanian, Université Paris Est - Créteil Val de Marne, France\\
Veronica Fernandez Marmol,	ITEFI (CSIC),	Spain	\\
Estrella Florido, Universidad de Granada,	Spain	\\
Vivian F. França,	São Paulo State University, Araraquara,	Brazil	\\
Sonja	Franke-Arnold,	University of Glasgow,	UK	\\
Pauline	Gagnon,	Indiana University,	USA	\\
Elisabetta	Gallo,	DESY,	Germany	\\
Gemma	García Alonso,	Universitat Autònoma de Barcelona,	Spain	\\
Ana	García Pietro,	Universidad del Pais Vasco,	Spain	\\
Carmen	García Recio,	Universidad de Granada,	Spain	\\
Laura	García-Álvarez,	Chalmers University of Technology,	Sweden	\\
Pascuala	García-Martínez,	Universitat de València,	Spain	\\
Erika Garutti, DESY, Germany	\\
Shohini	Ghose,	Wilfrid Laurier University,	Canada	\\
Elisabeth	Giacobino,	Laboratoire Kastler Brossel, École Normale Supérieure Paris,	France	\\
Ilaria	Gianani,	Università di Roma 3,	Italy	\\
Noel	Goddard,	Qunnect,	USA	\\
Fabienne	Goldfarb,	Université Paris-Saclay,	France	\\
Gabriela	González,	Louisiana State University,	USA	\\
María Carmen 	Gordillo,	Universidad Pablo de Olavide,	Spain	\\
Jimena D. 	Gorfinkiel,	The Open University,	UK	\\
Rachel	Grange,	ETH Zürich,	Switzerland	\\
Saïda 	Guellati-Khélifa,	Laboratoire Kastler Brossel Paris,	France	\\
Suchi	Guha,	University of Missouri, Columbia,	USA	\\
Montserrat  Guilleumas,	Universitat de Barcelona,	Spain	\\
Lucia	Hackermueller,	University of Nottingham,	UK	\\
Marta I.	Hernández,	Instituto de Física Fundamental IFF-CSIC,	Spain	\\
María Angeles	Hernández Vozmediano,	Institute of Materials Science of Madrid (ICMM), CSIC,	Spain	\\
Zoë 	Holmes,	EPFL, Lausanne,	Switzerland	\\
Yanhua	Hong,	Bangor University,	UK	\\
Ottavia	Jedrkiewicz,	Università dell'Insubria,	Italy	\\
Stacey	Jefferey,	CWI Amsterdam and Qsoft,	Netherlands	\\
Malgosia	Kaczmarek,	University of Southampton	,	UK	\\
Delaram	Kahrobaei,	The City University of New York,	USA	\\
Archana	Kamal,	Northwestern University,	USA	\\
Elham	Kashefi, NQCC, CNRS (LIP6) and University of Edinburgh,	France and UK	\\
Ursula	Keller,	ETH Zürich,	Switzerland	\\
Viv 	Kendon,	Strathclyde University,	UK	\\
Shelby	Kimmel,	Middlebury College,	USA	\\
Sabine	Klapp, Technische Universität Berlin, Germany	\\
Martina	Knoop, CNRS, Aix-Marseille Université, France\\
Natalia	Korolkova, University of St Andrews,	UK	\\
Svetlana	Kotochigova, Temple University,	USA	\\
Elica	Kyoseva,	NVIDIA Corp.,	USA	\\
Anne	L’Huillier,	Lund University,	Sweden	\\
Franziska	Lautenschläger,	Universität des Saarlandes,	Germany	\\
Cristina	Lazzeroni,	University of Birmingham,	UK	\\
Hanna	Le Jeannic,	CNRS -Laboratoire Kastler Brossel,	France	\\
Michèle	Leduc,	Laboratoire Kastler Brossel, École Normale Supérieure Paris,	France	\\
Inmaculada	Leyva,	Universidad Rey Juan Carlos,	Spain	\\
Ute	Lisenfeld,	Universidad de Granada,	Spain	\\
Maria Antonietta Loi,	University of Groningen,	Netherlands	\\
Rosa	Lopez, IFISC (UIB-CSIC),	Spain	\\
Sheila	López Rosa,	Universidad de Sevilla,	Spain	\\
Kathy	Lüdge,	Technische Universität Ilmenau,	Germany	\\
Agnes	Maitre,	Sorbonne Université- INSP,	France	\\
Julia	Maldonado-Valderrama,	Universidad de Granada,	Spain	\\
Laura	Mančinska,	University of Copenhagen,	UK	\\
Maria	Maragkou,	Riverlane,	UK	\\
Francesca Maria	Marchetti,	Universidad Autonoma de Madrid,	Spain	\\
Danijela Markovic,	CNRS/Thales,	France	\\
Irene	Marzoli, University of Camerino,	Italy	\\
Yvonne P. Mascarenhas, University of São Paulo, São Carlos,	Brazil	\\
Cristina Masoller,	Universitat Politècnica de Catalunya,	Spain	\\
Benedetta Mennucci, University of Pisa, Italy\\
Pérola	Milman, Laboratoire Matériaux et Phénomènes Quantiques, CNRS,  France	\\
Anna	Minguzzi, Laboratoire de Physique et Modélisation des Milieux Condensés, France	\\
Elisa	Molinari,	University of Modena e Reggio Emilia,	Italy	\\
Juliette Monsel,	Chalmers University of Technology,	Sweden	\\
Arianna	Montorsi,	Politecnico di Torino,	Italy	\\
Giovanna Morigi,	Universität des Saarlandes,	Germany	\\
Silvia	Motti,	University of Southampton,	UK	\\
Olga	Muñoz,	Instituto de Astrofísica de Andalucía, CSIC,	Spain	\\
Elke	Neu-Ruffing,	RPTU Kaiserslautern,	Germany	\\
Síle Nic Chormaic, Okinawa Institute of Science and Technology, Japan \\
Beatriz	Olmos Sanchez,	Eberhard-Karls-Universität Tübingen, Germany	\\
Alicia	Palacios,	Universidad Autónoma de Madrid, Spain	\\
Carmen Palacios Berraquero, Nu-Quantum,	UK	\\
Elisa Palacios Lidon, Universidad de Murcia,	Spain	\\
Adriana Pálffy-Buß, Julius-Maximilians-Universität Würzburg, Germany\\
Nancy	Paul,	CNRS, Laboratoire Kastler Brossel,	France	\\
Hélène	Perrin,	Laboratoire de Physique de Lasers, CNRS, 	France\\
Isabelle Philip, University of Montpellier,	France	\\
Silvina	Ponce Dawson,	Universidad de Buenos Aires and CONICET, Argentina	\\
Natacha	Portier, ENS Lyon,	France	\\
Àngels	Ramos,	Universitat de Barcelona,	Spain	\\
Chitra	Rangan,	University of Windsor, Canada\\
Stephanie Reich, Freie Universität Berlin,	Germany	\\
Lucia	Reining, Laboratoire des Solides Irradiés, École Polytechnique, France	\\
Rebeca	Ribeiro Palau,	CNRS, Centre de nanosciences et de nanotechnologies, France	\\
Monika	Ritsch-Marte,	Medizinische Universität Innsbruck, Austria	\\
Isabelle Robert-Philip,	Laboratoire Charles Coulomb - CNRS,	France	\\
Chiara	Roda, Department of Physics, University of Pisa and INFN-Pisa, Italy	\\
Jacqui	Romero,	University of Queensland,	Australia	\\
Halina	Rubinsztein-Dunlop,	The University of Queensland,	Australia	\\
Berta	Rubio Barroso,	Instituto de Física Corpuscular, CSIC-Uni. Valencia, Spain	\\
Gemma Ruis,	IMB-CNM-CSIC, Spain\\
Ana Belen Sainz,	University of Gdansk,	Poland	\\
Catherine Schwob,	Sorbonne Université- INSP,	France	\\
Signe 	Seidelin,	Université Grenoble Alpes,	France	\\
Irene	Sendiña Nadal, 	Universidad Rey Juan Carlos, 	Spain	\\
Efrat Shimshoni, Bar-Ilan University,	Israel	\\
Ruth Signorell,	ETH Zürich,	Switzerland	\\
Sudeshna Sinha,	Indian Institute of Science Education and Research Mohali, India	\\
Clivia	Sotomayor Torres, International Iberian Nanotechnology Laboratory, Portugal	\\
Janine	Splettstoesser,	Chalmers University of Technology, Sweden	\\
Magdalena Stobinska, University of Warsaw, Poland\\
Giovanna Tancredi, Chalmers University of Technology,	Sweden	\\
Marika	Taylor,	University of Birmingham,	UK	\\
Sarah	Thomas,	Imperial College London,	UK	\\
Angela	Thränhardt,	Technische Universität Chemnitz,	Germany	\\
Giovanna	Tissoni,	Institut de Physique de Nice,	France	\\
Laura	Tolos, Institute of Space Sciences (ICE-CSIC),	Spain	\\
Silvia	Torres-Peimbert,	Universidad Nacional Autónoma de México,	Mexico	\\
Rosa	Tualle-Brouri,	Institut d’Optique Graduate School,	France	\\
Elinor	Twyeffort,	University of Southampton,	UK	\\
Anu	Unnikrishnan,	ETH Zürich,	Switzerland	\\
Joan Vaccaro, Griffith University, Australia	\\
Roser Valenti, Goethe-Universität Frankfurt,	Germany	\\
Licia Verde, University of Barcelona, Spain	\\
Francesca Vidotto, University of Western Ontario,	Canada	\\
Patrizia Vignolo, Institut de Physique de Nice,	France	\\
Silvia	Vignolo, Max Planck Institute of Colloids and Interfaces, Potsdam, Germany	\\
Lorenza	Viola,	Dartmouth College,	USA	\\
Smitha	Vishveshwara,	University of Illinois Urbana Champaign	,	USA	\\
Valia	Voliotis,	Sorbonne Université, INSP,	France	\\
Lucia	Votano,	Istituto Nazionale di Fisica Nucleare,	Italy	\\
Stephanie Wehner,	Delft University of Technology,	Netherlands	\\
Carrie	Weidner,	University of Bristol,	UK	\\
Ulrike	Woggon,	TU Berlin,	Germany	\\
Delphine	Wolfersberger, Laboratoire Matériaux Optiques, Photonique et Systèmes, CentraleSupélec, France	\\
Susanne	Yelin, Harvard University, USA	\\
Pauline	Yzombard, Sorbonne Université, Laboratoire Kastler Brossel, France	\\
Ilaria	Zardo,	Universität Basel,	Switzerland	\\
Almudena Zurita, Universidad de Granada,	Spain	\\
Magdalena Zych,	Stockholm University,	Sweden	\\


\end{document}